Investigating High School and Pre-High School Teachers' Perceptions and Experiences Introducing Quantum Concepts: A Survey of QuanTime and other Quantum-related Activities

Apekshya Ghimire[1], Jaya Shivangani Kashyap[1], Emily Edwards[2], Diana Franklin[3] and Chandralekha Singh[1]

[1]University of Pittsburgh, Department of Physics and Astronomy, Pittsburgh, PA 15260

[2]Duke University, Department of Electrical and Computer Engineering, Durham, NC 27708

[3]University of Chicago, Department of Computer Science, Chicago, IL 60637

**Abstract**

This study investigates the experiences of pre-high and high school teachers in implementing QuanTime and other quantum-related activities aiming to promote quantum literacy and introduce foundational quantum concepts to K-12 students. The ultimate goal is to help prepare a diverse future workforce in quantum information science and technology (QIST). Teachers were divided into two groups: pre-high school (grades 4-8) and high school (grades 9-12). We used a survey featuring 12 Likert-scale questions and 14 open-ended responses to assess teachers' perceptions, engagement, and feedback about engaging in QuanTime and other quantum-related activities. Approximately two-thirds of the teachers responding to the survey implemented QuanTime activities in their classes. High school teachers who responded to the survey were most likely to use activities like Wave-Particle Duality and Electron Transitions while pre-high school teachers showed a strong interest in Art & Polarization. Open-ended feedback highlighted the ease of integrating these activities into existing curricula and the minimal preparation required, making them accessible for educators. The positive reception across both groups indicates that QuanTime and other quantum-related activities are valuable tools for early-age quantum education. By engaging students with quantum concepts from a young age, these activities have the potential to spark interest, which may contribute to their future engagement over time. It can inspire a diverse group of students and has the potential to get them interested in future opportunities in the growing field of QIST.

# 1. Introduction

We are currently living in the era of the second quantum revolution [1], a time characterized by rapid advancements in quantum technologies that have the potential to transform the world and reshape our understanding of science and technology [2, 3]. Unlike the first quantum revolution [4], which primarily focused on understanding the principles of quantum mechanics to develop transistors and lasers as well as harness nuclear energy, etc., the second quantum revolution leverages principles like superposition and entanglement to develop applications such as quantum computers, quantum cryptography, and quantum internet [5-8]. The potential of these technologies is beginning to gain recognition across a variety of sectors, from national security, finance, and healthcare to communications and materials science. Additionally, novel quantum technologies of the 21st century promise to enhance basic science research. As a result, there is an increasing demand for professionals who are equipped with the knowledge and skills needed to work in quantum-related fields [9-13].

It is essential to raise awareness and understanding of quantum concepts among a broader audience to fully harness the opportunities presented by this quantum information revolution [14-17]. Quantum literacy is becoming a critical component of modern education as industries look to build a workforce capable of navigating and utilizing emerging quantum technologies [9, 12, 18]. In this context, quantum literacy refers to enabling students at the K–12 level to begin learning about quantum science and technology, laying the foundation for future engagement with quantum ideas. Promoting quantum literacy is not only important for individuals who aspire to become quantum scientists or engineers; it is equally vital for the public, who can benefit from a basic understanding of the field [15, 16, 19]. A knowledge of foundational quantum concepts can help individuals make informed decisions in a world where quantum technologies may soon influence different aspects of life, from secure communication systems to enhanced data processing capabilities. Furthermore, this knowledge can open doors to quantum-adjacent jobs [15, 16, 19-21], roles that do not require deep expertise in quantum mechanics but still demand a familiarity with the basic principles of the field, such as positions in quantum software development, quantum policy, or quantum communication.

Despite the growing recognition of the importance of quantum literacy [22-24], there is currently no standardized approach to teaching quantum concepts in the K-12 curriculum in the United States [25-27], particularly within the framework of the Next Generation Science Standards (NGSS) [28, 29]. The NGSS, widely adopted across the United States to guide science education, does not prescribe specific methods or content for introducing students to quantum mechanics except for the standard curriculum that includes atoms and periodic table [30]. This lack of guidance leaves many educators underprepared and uncertain about how to effectively incorporate these topics into their lessons [25]. The absence of a structured curriculum on quantum science represents a significant missed opportunity to introduce students to these revolutionary concepts at an early age, potentially sparking an interest in STEM (Science, Technology, Engineering, and Mathematics) subjects and preparing them for future careers in these domains [31]. This is in contrast to publicly available curricular information in Europe where quantum topics are integrated into K-12 science curricula, and there is evidence to suggest that teaching quantum science is a vehicle for students to engage with ideas of the nature of science [32-35].

This is where innovative initiatives focused on early quantum education like QuanTime [36] come into play. Inspired by the success of the "Hour of Code" initiative [37], which has introduced millions of students to the basics of coding through engaging and interactive activities, QuanTime is designed as a pre-packaged, interactive resource that helps K-12 teachers introduce quantum concepts to their students in a fun and approachable manner. QuanTime is a program that was created and currently managed under the National Q-12 Education Partnership. The name "QuanTime" encapsulates the idea of dedicating time specifically for exploring quantum science, making it as accessible and enjoyable as the Hour of Code does for computer programming. The activities curated and disseminated by QuanTime are tailored to fit within existing science lessons, offering adaptable tools that teachers can seamlessly integrate, even in the absence of a defined quantum syllabus. The materials of QuanTime are designed so that the teachers do not require additional time for preparation nor additional support for implementation. The activities target foundational quantum topics such as wave-particle duality, electron transitions, quantum entanglement, quantum art and polarization, and different types of quantum games. Teachers are encouraged to select activities that align with their grade level and subject area. The activities include background materials and implementation guides to support teachers with limited prior experience in quantum physics. By providing these ready-to-use resources, QuanTime has the potential to empower educators to introduce complex quantum topics in a way that is understandable and engaging for young learners.

There has been significant work in developing engaging and motivating approaches to introduce students to foundational quantum concepts. For example, Hahn and Gire used physical movements to help students grasp basic ideas [38], while Lopez-Incera et al. designed interactive games where students act as particles and scientists, simulating real lab experiences to explore quantum information science and technology (QIST) [39, 40]. These activities help students build mental models of quantum concepts, including probabilistic measurement, entanglement, decoherence, and quantum cryptography. Expanding on these game-based methods, Marckwordt et al. introduced a dodgeball-inspired game to teach quantum entanglement [41]. Other instructional strategies include comparing classical and quantum two-state systems [42], using diagrammatic approaches and incorporating basic mathematics [43]. In secondary education, various researchers have developed teaching-learning sequences, inquiry-based activities, and experiential programs to enhance students' understanding of quantum mechanics. Efforts such as those by Michelini et al. [44], Bitzenbauer [45], and Stadermann et al. [46] focus on integrating quantum physics into school curricula through innovative materials, inquiry-driven methods, and connections to the nature of science, making quantum concepts more accessible and engaging for students. QuanTime builds upon these prior efforts by offering structured, adaptable activities that align with teachers' needs such as limited classroom time, varying levels of familiarity with quantum content, and the need for standards-aligned, ready-to-use materials that can be easily integrated into existing science curricula and classroom settings. By incorporating interactive and inquiry-driven learning approaches, QuanTime provides students with engaging experiences that reinforce fundamental quantum concepts and increase their motivation to learn.

Introducing young students to quantum concepts through exciting and fun activities has several benefits [47-50]. First, it helps to demystify the often-intimidating subject of quantum mechanics by presenting it in a simplified and engaging manner. By exposing students to these ideas early on, we can ignite a curiosity and interest in quantum science, potentially attracting a more diverse group of students to the field [51, 52]. This early exposure is critical for diversifying the future workforce in QIST, which has traditionally been dominated by individuals from specific educational and demographic backgrounds [53]. Increasing diversity in quantum-related fields can lead to a wider range of perspectives and innovations, driving the field forward in novel and unexpected ways.

Moreover, familiarizing students with the fundamental differences between classical mechanics and quantum mechanics at a young age can have lasting educational benefits [54, 55]. Quantum mechanics, with its counterintuitive concepts like superposition and entanglement [56], is often challenging for students to grasp, particularly if they encounter it for the first time at the undergraduate or graduate level. By laying the groundwork for these concepts early in a student's educational journey, we can help make the subject less daunting and easier to comprehend later. This foundational knowledge can serve as a springboard for deeper learning, allowing students to build on their understanding of quantum mechanics as they progress in their academic careers.

We drew upon two related theoretical frameworks: Expectancy-Value Theory (EVT) [57] and Social Cognitive Career Theory (SCCT) [58] to conceptually ground the broader goals of this research and development effort aimed at enhancing K–12 student engagement with quantum concepts. The EVT is a framework that emphasizes that students are more motivated to learn when they believe they can succeed (expectancy) and see value in the subject they are learning [57]. If K-12 students have confidence that they can understand quantum concepts and they can see how quantum concepts relate to their future careers and daily life by developing a broader scientific understanding early on, they are more likely to stay engaged even after completing their K-12 education. In particular, if they find quantum science and technology interesting or relevant, they are more likely to invest time and effort in learning it. The SCCT is a framework that supports EVT by considering how confidence, expectations about future outcomes, and environmental factors influence students' academic interests and career choices [58]. Introducing students to quantum concepts early through QuanTime and other quantum-related activities can boost their confidence in STEM and spark interest in quantum-related careers by showing real-world applications of quantum science. While these frameworks did not directly inform the design of the QuanTime activities or the specific survey items, they broadly helped frame the underlying rationale for introducing quantum science to K–12 students and investigating teachers' perspectives on these efforts.

Aligned with these frameworks, Hidi and Renninger's four-phase model of interest development [59] outlines how situational interest, sparked by meaningful and engaging experiences, can evolve into a well-developed individual interest over time. A major goal of introducing QuanTime and other quantum-related activities at the K–12 level is to intentionally spark such situational interest in quantum science early on, with the hope that repeated exposure and positive experiences will nurture a more sustained and self-driven engagement with the subject.

By making quantum concepts accessible and relevant, these activities aim to support students' confidence in their ability to understand quantum topics (expectancy), help them recognize both intrinsic value (e.g., curiosity, enjoyment) and extrinsic value (e.g., STEM career pathways), and ultimately influence their long-term motivation and aspirations. In doing so, these efforts are not only striving to foster immediate interest and engagement but also lay the foundation for students to see themselves as capable of and interested in pursuing quantum science as part of their academic and professional futures.

These frameworks serve as an overarching lens through which we interpret and contextualize our findings. Our study focuses on how high school and pre-high school teachers view the importance of QuanTime and other quantum-related activities and their influence on students. For example, we want to understand whether teachers who use them view QuanTime and other quantum-related activities as helping students stay motivated, engaged, and interested in quantum science. Through these early experiences with quantum-related activities, our goal is to support and promote the development of a diverse and well-prepared workforce that can contribute to the future of quantum science and technology.

To evaluate the impact of QuanTime and other quantum-related activities on teacher experiences and student engagement with quantum concepts, we conducted a survey with both pre-high school and high school teachers. The participants were divided into two groups: pre-high school teachers (covering grades 4-8) and high school teachers (covering grades 9-12). We surveyed these teachers using a combination of 12 Likert scale questions and open-ended response questions which were validated to gather insights regarding their perceptions of quantum concepts, their experiences with implementing QuanTime and other quantum-related activities, and the impact on student interest and engagement. The survey aimed to gather teachers' self-reported confidence in teaching quantum concepts and their perspectives on incorporating quantum-related activities into their instruction, the level of support they felt they needed from quantum experts beyond what was provided in the packaged QuanTime or other quantum-related activities, and their observations of student reactions to learning about quantum concepts.

By analyzing the data, we aimed to understand the experiences of the pre-high school and high school teachers in implementing QuanTime and other quantum-related activities in their classrooms and whether they could serve as a practical tool for introducing quantum concepts in K-12 education. The feedback from the teachers who specifically used QuanTime activities provided valuable insights into the strengths and areas for improvement of the QuanTime initiative, informing future efforts to refine and expand the program. This study contributes to the growing body of research focused on innovative approaches to STEM education and highlights the importance of integrating quantum literacy into early education to prepare a diverse, well-informed workforce for the challenges and opportunities of the quantum era.

## 2. Methodology

### 2.1 Survey Design

To evaluate the perception of the K-12 teachers on teaching quantum concepts and implementing QuanTime and other quantum-related activities, a preliminary version of the survey was developed by the researchers using an iterative process. Then, after the researchers were satisfied with all the important issues they wanted to investigate in the survey and the wording of each question, we validated the survey by iterating it with six PhD students specializing in physics education, three physics professors, and six K-12 teachers and/or educational leaders at K-12 level. Once the survey was in its final form, we reached out to middle school and high school teachers who signed up to receive information through the National Q-12 Education partnership program for their responses to the survey focused on QuanTime and other quantum-related concepts informing them that this was a research project.

As noted, the goal of this survey was to gather teachers' perspectives on introducing quantum concepts and implementing QuanTime and other quantum-related activities in their classrooms. The survey aimed to gather insights into teachers' confidence in organizing these quantum-related activities, their needs for time or support, student engagement and enjoyment, and whether students expressed further interest in quantum concepts or related careers. To include any K-12 teachers who wanted to provide feedback about quantum-related activities and ensure a comprehensive understanding of related issues, the survey questions were designed to learn about both the implementation of QuanTime and other quantum-related activities more broadly regardless of whether the teachers incorporated QuanTime activities in their classes. In particular, teachers who completed the survey were not required to have conducted a QuanTime activity; they could have been teaching any quantum concepts and conducting quantum-related activities with their students. Moreover, teachers who implemented QuanTime activities may have also taught additional quantum concepts, which were also investigated in the survey.

The survey consisted of 26 questions, including 12 on a Likert scale from 1 to 7 (1-Strongly Disagree, 2-Slightly Disagree, 3-Disagree, 4-Neutral, 5-Slightly Agree, 6-Agree, 7-Strongly Agree), while the rest were open-ended. These open-ended questions covered topics like the subjects the teachers taught, the number of students involved in quantum activities, the grade levels in which QuanTime or other quantum-related activities were implemented, the specific activities used to teach quantum concepts, and whether the activities were linked to the curriculum (or a fun addition). We also asked teachers to identify the most exciting aspects of quantum learning for their students, share any challenging questions their students asked them, and provide suggestions and feedback on the QuanTime and other quantum-related activities. Teachers were also asked to specify the type of school in which they teach (e.g., public, charter, private, homeschool, parochial, etc.)

## 2.2 Participants

For this study, we reached out to 132 teachers through our Q-12 partnership program. These teachers had expressed interest in quantum-related activities on the Q-12 partnership website and had signed up to receive potentially relevant information. As a result, the survey was distributed specifically to this group of self-selected teacher group. Of those, 40 teachers participated in the survey, representing a range of Science, Technology, Engineering, and Mathematics (STEM) subjects across different grade levels. We discarded four responses that involved completing less than 60% of the survey. Ultimately, 36 teachers including one outreach coordinator at an after-school program completed more than 60% of the survey, providing a diverse set of insights into their experiences engaging students with QuanTime and other quantum-related activities. The survey responders represented a broad range of STEM disciplines, including physics, chemistry, mathematics, earth and life sciences, engineering, technology, computer science, and environmental science. We have also provided the distribution of the high school teachers by subject area in Table 1. The majority of teachers reported teaching in public schools, while a smaller number indicated private schools, and one mentioned homeschooling. However, we did not gather information regarding their academic backgrounds or their preparedness to teach quantum physics. It is unclear why the other teachers did not participate in the survey, and we do not know whether they implemented QuanTime or other quantum-related activities. As a result, we have no insight into their perspectives on these activities.

Table 1. Distribution of 23 high school teachers by subject area

| Subject Area | Number of Teachers |
| --- | --- |
| Physics | 7 |
| Chemistry | 6 |
| Physics and Chemistry | 6 |
| Others (Computer Science, Engineering, World Language, etc.) | 4 |

We divided the survey respondents into two groups based on the grade levels they taught. Teachers who taught classes from $4^{th}$ to $8^{th}$ grades were grouped as pre-high school teachers, while those who taught grades 9 through 12 were categorized as high school teachers. There was one teacher/staff member at an after-school program providing informal learning, who noted that they provided QuanTime instruction to their staff after going through some of the activities themselves. This individual is the same person referred earlier as an "outreach coordinator". Out of the 36 respondents who completed the survey, 12 were pre-high school (Pre-HS) teachers teaching grades 4 through 8, while 23 were high school (HS) teachers teaching grades 9 through 12. This distinction between Pre-HS and HS levels allowed us to compare the responses of teachers from different educational levels and gain insights into any variations in perceptions or engagement with the QuanTime or other quantum-related activities. We divided the teachers into two groups to align with the structural design of the QuanTime program and the differences in how quantum concepts are introduced at each level. While the overarching goal of sparking curiosity and developing interest in quantum science applies to both groups, the conceptual depth and instructional approach of the activities vary. The National Q-12 Partnership [36], which coordinates QuanTime, provides separate sets of activities designed specifically for pre-high school and high school students, considering their developmental readiness and curriculum context. Activities for younger students are generally more exploratory and visual, focusing on intuitive engagement, while those for older students incorporate deeper conceptual elements to help them make connections to broader scientific ideas. Additionally, many teachers and teacher leaders focus on a specific educational level, so this division allows for clearer interpretation of the findings and more targeted insights for practitioners.

### 2.3 Survey content and Data Analysis

As noted, the quantitative data were gathered using a Likert scale, with 12 questions rated on a scale from 1 to 7, where 1 represented "strongly disagree" and 7 represented "strongly agree". This allowed us to measure the teachers' level of agreement or disagreement on various aspects of the activities, such as their confidence in organizing the quantum activities, student engagement, and the relevance of the activities to their curriculum. The Likert-scale responses were analyzed quantitatively to determine average ratings for engagement, relevance, and student understanding. Even though the responses are based on a 7-point Likert scale, we have grouped them into three main categories to simplify the analysis: Agree, Disagree, and Neutral. "Agree" includes both "Agree" and "Strongly Agree", "Disagree" combines "Disagree" and "Strongly Disagree" and "Neutral" encompasses "Slightly Disagree", "Slightly Agree" and "Neutral".

In addition to the Likert scale, the survey contained open-ended questions that enabled teachers to provide qualitative feedback. A qualitative analysis of open-ended responses to the questions was conducted to identify teachers' views regarding student reactions, challenges faced by the teachers, and their overall feedback on the QuanTime and other quantum-related activities. These responses were structurally coded [60]. Structural coding relates to "a specific research question used to frame the interview" [60, 61]. Consistent with the structural coding [60], which is a holistic coding approach, we kept teachers' comments for each of the questions as a category characterized by the question they were answering. Moreover, the combination of quantitative and qualitative data provided insight into teachers' experiences with implementing QuanTime and other quantum-related activities, as well as their perceptions of student engagement.

### 2.4 Quantum Activities Overview

In our sample, 9 of the 12 pre-high school teachers implemented the QuanTime activity in their classes, and 16 of the 23 high school teachers did the same. Thus, including Pre-HS and HS classes, 25 conducted the QuanTime Activity in their classes out of the 36 teachers who completed the survey and had their students engage in any quantum-related activities. This implies that approximately two-thirds of the teachers selected QuanTime activities. The rest of the teachers implemented other quantum-related activities of their choice from other sources. These activities included introductory quantum computing concepts, diffraction with LEDs, lasers, quantum cryptography, spectroscopy, among others. One teacher specifically mentioned using resources provided by Quantum for All [17, 62]. However, we do not have detailed information on the specific sources of the activities used by the other teachers or their reasons for selecting those particular activities, as these aspects were not asked for in the survey.

The QuanTime activities included a range of hands-on, interactive lessons designed to introduce fundamental quantum concepts to K-12 students in a packaged form so that the teachers could use them easily. All activities were intended for facilitators without prior quantum knowledge, focused on minimal learning goals, and were not part of a structured curriculum designed to build knowledge progressively over time. Activities such as "Art & Polarization", "Electron Transitions" and "Wave-Particle Duality" were among the most frequently mentioned by the teachers. These activities are intended to engage students in practical, visual demonstrations that make abstract quantum concepts more accessible.

We note that the goal of the survey was to learn about teachers' perceptions and engagement with quantum activities in pre-college classrooms. However, QuanTime was specifically mentioned in the survey questions because we had promoted these activities through the Q-12 partnership program. Therefore, our results are divided into three parts. First, we discuss all of the quantitative results combining QuanTime and other quantum-related activities for teachers at high school and pre-high school levels. Then, we focus specifically on the QuanTime activities that teachers did in their classes. Finally, we present teachers' responses to the open-ended questions, highlighting their reflections, experiences, and suggestions regarding all the activities. While QuanTime activities were designed to include clear instructions and align with existing science curricula, the survey data did not reveal any noticeable differences in how teachers responded based on whether they used QuanTime or other quantum-related activities. The time spent on QuanTime activities also varied considerably across classrooms, similar to the variation observed with other quantum-related activities that teachers selected on their own. Given this variability, our data do not suggest any qualitative differences in teacher responses between the two types of activities.

Teachers implemented these activities across a variety of subjects and grade levels. The amount of time spent on each activity varied, with some teachers incorporating them into a single class period, while others using them as part of extended lessons or units.

### 3. Results and Discussion

**3.1 Teacher Engagement with QuanTime and Other Quantum Related Activities**

As noted, approximately two-thirds of the teachers conducted the QuanTime Activity while one-third helped their students engage with quantum concepts using activities they obtained from other sources. In this section, we discuss the effectiveness of all these quantum-related activities overall (regardless of whether they were QuanTime or other quantum-related activities) based upon teachers' views about whether and how much they helped students learn about quantum concepts or develop an interest in learning about quantum concepts and careers. Thus, our primary focus here is on understanding teacher engagement and their overall feedback pertaining to the activities they conducted.

We analyzed responses to 12 questions on a Likert scale ranging from 1 to 7 (a score of 4 signifies a neutral response). As noted, we categorized the teachers into two distinct groups: pre-high school teachers (covering K-8) and high school teachers (covering grades 9-12). The average scores were calculated for each statement across all participating teachers as shown in Table 2. By comparing the average scores of these two groups, we aimed to identify any practically significant differences in their responses. This approach allowed us to explore whether the engagement, perceptions, and attitudes varied between teachers at different educational levels regarding these quantum activities. This type of analysis also provided insights into potential differences in educational approaches, needs, and familiarity with the content across different teaching levels.

Table 2: Comparison of average Likert scale scores for Pre-High School (Pre-HS) and High School (HS) Teachers Across 12 Statements (1-Strongly Disagree, 2-Slightly Disagree, 3-Disagree, 4-Neutral, 5-Slightly Agree, 6-Agree, 7-Strongly Agree).

| Survey Questions | All | HS | Pre-HS |
| --- | --- | --- | --- |
| 1. I feel that students should learn quantum concepts in high school, middle school or earlier | 6.0 | 6.1 | 5.7 |
| 2. I feel that QuanTime activities/teaching quantum concepts fit well with the curriculum I am using with my students | 5.5 | 5.7 | 5.2 |
| 3. I felt confident using QuanTime activities/teaching quantum concepts to my students | 5.7 | 5.9 | 5.4 |
| 4. My school administrators are flexible and supportive of my incorporating QuanTime activities/teaching quantum concepts to my students | 5.6 | 5.8 | 5.2 |
| 5. My students enjoyed QuanTime activities/learning quantum concepts | 5.8 | 6.1 | 5.3 |
| 6. Most of my students were engaged in QuanTime activities/learning quantum concepts | 5.7 | 6.0 | 5.2 |
| 7. My students expressed interest in quantum-related careers after doing QuanTime activities/learning quantum concepts | 4.7 | 4.9 | 4.3 |
| 8. My students expressed interest in learning more about quantum concepts in the future after doing QuanTime activities/learning quantum concepts in my class | 5.2 | 5.4 | 4.8 |
| 9. The amount of time I spent on QuanTime activities/learning quantum concepts was sufficient to complete the unit in my class | 5.0 | 5.2 | 4.7 |

| 10. I needed more preparation time to conduct the QuanTime activities/teach quantum concepts well | 4.0 | 3.9 | 4.0 |
|---|---|---|---|
| 11. I would have liked more support during the implementation of the QuanTime activities/teaching of quantum concepts | 3.6 | 3.8 | 3.3 |
| 12. I would like to do the QuanTime activities/teach quantum again next year | 6.2 | 6.3 | 5.9 |

From Table 2, we observe that the average Likert scale scores tend to be higher for high school teachers across most statements, except for Statement 10. Most of these statements have average scores greater than 4 for both pre-high school and high school teachers, indicating that the teachers largely agreed with these statements. However, Statements 10 and 11 have lower average scores compared to other questions (either 4 or somewhat lower than 4, which is neutral). Notably, these statements pertain to the perceived need for additional preparation time and support for engaging students with the QuanTime and other quantum-related activities. Regardless of the educational level, teachers did not feel that they required significant extra preparation time or support for the implementation of these quantum activities. This finding suggests that the pre-college quantum-related activities may be effectively designed to be teacher-friendly and easily integrated into classroom settings, making teachers feel confident and comfortable in utilizing them without needing extensive additional resources or preparation.

The following pie charts (Fig. 1-12) provide a detailed breakdown of how teachers responded to each statement, showcasing the variations in agreement or disagreement across the Likert scale (1-7). For this analysis also, we combined the responses from all the teachers to analyze data in the results presented below, including those who engaged students with QuanTime or other quantum-related activities. Moreover, the responses for all activities have been aggregated rather than separated by individual activity, providing an overall view of teacher perspectives. These visualizations help to illustrate the overall trends and specific differences in perceptions between the two groups for each statement, offering deeper insights into teachers' views on various aspects of the QuanTime or other quantum-related activities.

Each pie chart shows the response distribution, moving clockwise from "Strongly Disagree" at the top to "Strongly Agree". Although the pie chart displays seven distinct categories, we grouped the responses into three broader categories: Agree, Disagree, and Neutral, for the purpose of analysis and discussion. If a response category is missing in a particular pie chart, it indicates that no teachers selected that option, meaning the percentage for that category is 0%. To enhance readability, we include only a few pie charts in the main text—specifically those where the difference in agreement ("Strongly Agree" and "Agree" combined) between pre-high school and high school teachers exceeds 10%. Pie charts with a difference of 10% or less are included in the Appendix to allow interested readers to explore the full range of teacher responses.

The first statement is, "I feel that students should learn quantum concepts in high school, middle school, or earlier". Fig. 7 (in Appendix) shows that for the high school level, a large majority of teachers either agreed or strongly agreed with the statement, totaling 83% support.

Similarly, at the pre-high school level, 76% of teachers either agreed or strongly agreed. This is likely because the teachers in our survey sample chose to voluntarily participate in quantum education activities. A small percentage remained neutral, while only one person expressed disagreement from both groups. Both groups of educators demonstrated considerable support for introducing quantum concepts at earlier educational stages, with slightly higher enthusiasm observed among high school teachers.

The second statement is, "I feel that QuanTime activities/teaching quantum concepts fit well with the curriculum I am using with my students". Fig. 1 shows that 78% of the high school teachers either agreed or strongly agreed with the statement. For the pre-high school level, 58% of teachers either agreed or strongly agreed, 34% of respondents remained neutral, and 8% strongly disagreed. The results indicate broad support among both groups of educators for integrating quantum concepts into their existing curricula, though enthusiasm appears significantly higher among high school teachers.

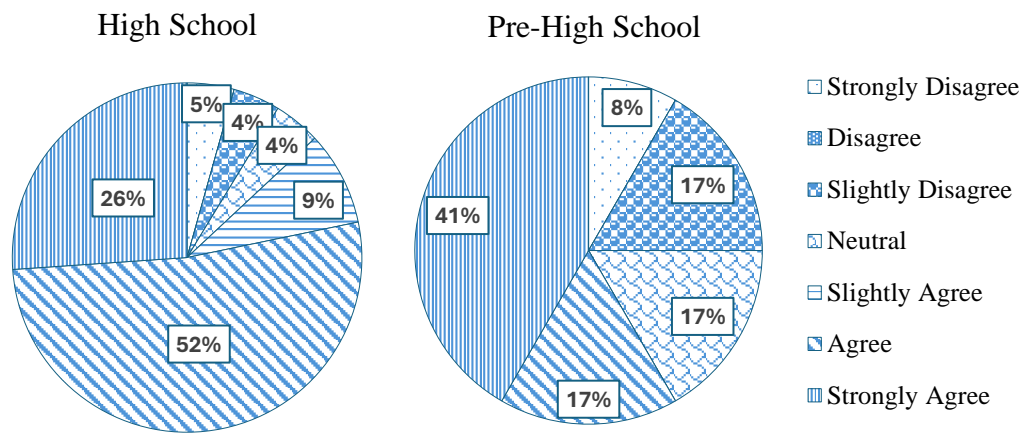

Fig. 1: Percentage responses of the high school and pre-high school teachers to Statement 2

The third statement is: "I felt confident using QuanTime activities/teaching quantum concepts to my students". Fig. 2 shows that among high school teachers, 81% either agreed or strongly agreed with this statement, while the remaining respondents were neutral (including slightly agree and slightly disagree). For pre-high school teachers, 59% either strongly agreed or agreed, 33% were neutral, and 8% (one teacher) disagreed. Overall, most teachers expressed confidence incorporating QuanTime and other quantum-related activities into their classes.

Statement 3

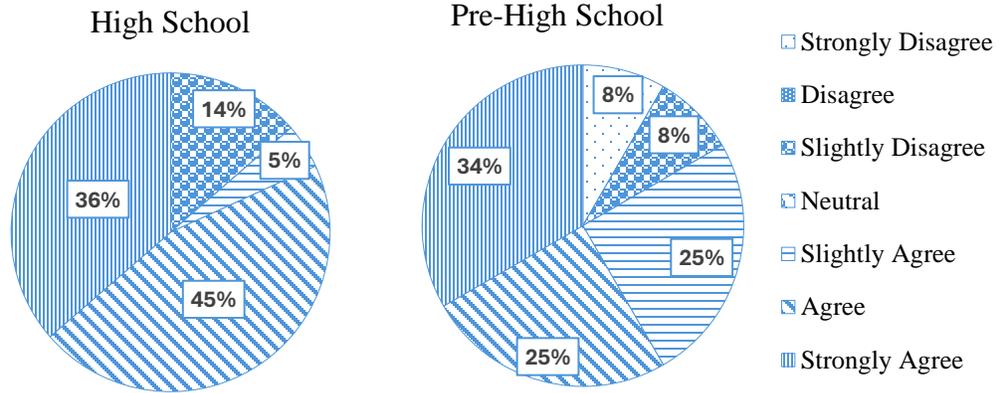

Fig. 2: Percentage responses of the high school and pre-high school teachers to Statement 3

The fourth statement is: "My school administrators are flexible and supportive of my incorporating QuanTime activities/teaching quantum concepts to my students". Fig. 3 shows that 74% of the high school teachers strongly agreed or agreed with this statement. 22% were neutral and 4% (1 teacher) disagreed. 59% of the pre-high school teachers strongly agreed or agreed while 25% were neutral and 16% disagreed. These differences highlight variations in perceived administrative support between educational levels.

Statement 4

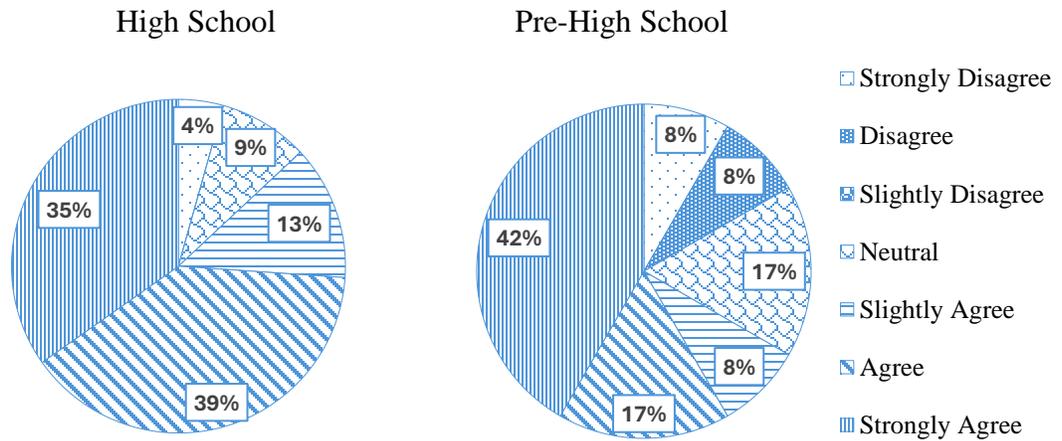

Fig. 3: Percentage responses of the high school and pre-high school teachers to Statement 4

The fifth statement is: "My students enjoyed QuanTime activities/learning quantum concepts." Fig. 4 shows that among high school teachers, 82% strongly agreed or agreed, indicating that the majority observed positive student engagement. Meanwhile, 18% were neutral. For pre-high school teachers, 68% strongly agreed or agreed with the statement, reflecting a significantly lower level of observed student enjoyment compared to high school level. 16% were neutral, while a notable 16% disagreed, suggesting mixed experiences with student reception at this level.

Statement 5

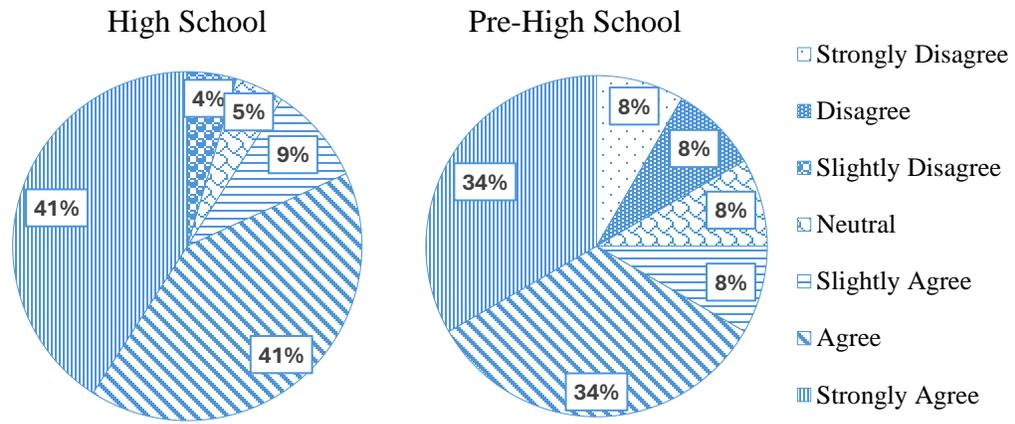

Fig. 4: Percentage responses of the high school and pre-high school teachers to Statement 5

The sixth statement is: "Most of my students were engaged in QuanTime activities/learning quantum concepts." Fig. 8 (in Appendix) shows that among high school teachers, 70% strongly agreed or agreed, indicating a majority found their students actively engaged in these activities, while 30% were neutral. For pre-high school teachers, 67% strongly agreed or agreed, showing slightly lower observed engagement compared to high school students. Meanwhile, 16% were neutral, and 17% disagreed, reflecting small variance in perceived student engagement within these levels.

The seventh statement reads: "My students expressed interest in quantum-related careers after doing QuanTime activities/learning quantum concepts." Fig. 9 (in Appendix) shows that at the high school level, only 26% agreed with this statement, while the majority (70%) were neutral, and 4% disagreed. For pre-high school teachers, a similar trend was observed, with 25% agreeing, 64% remaining neutral, and 9% disagreeing. These results suggest that while some teachers felt that students showed an interest in quantum-related careers, most either remained indifferent or thought that their students did not have a clear inclination toward such careers after engaging with the activities. It is unclear whether the teachers who agreed with this statement did so based on interest expressed by a single student or by multiple students.

The eighth statement reads: "My students expressed interest in learning more about quantum concepts in the future after doing QuanTime activities/learning quantum concepts in my class." Fig. 10 (in Appendix) shows that among high school teachers, 56% agreed with this statement, while the remaining teachers were neutral. For pre-high school teachers, 50% agreed, 41% were neutral and 9% disagreed. This suggests that a moderate portion of both groups observed an increased interest in quantum concepts among their students, although many remained neutral or undecided. The relatively high number of neutral responses, along with a small but notable portion of disagreement, suggests that many teachers felt that their students may not have developed a strong or lasting interest in quantum concepts, despite engaging in QuanTime or other quantum-related activities. This finding may indicate a diversity in student interest to further their understanding, a need for further refinement in how quantum concepts are presented to make more students excited or more support in fostering deeper student interest in learning quantum concepts.

The ninth statement is: "The amount of time I spent on QuanTime activities/learning quantum concepts was sufficient to complete the unit in my class". Fig. 5 shows that among high school teachers, 52% agreed that the time allocated was sufficient, 44% were neutral and 4% disagreed. For pre-high school teachers, only 33% agreed, and a significant portion (59%) remained neutral, while 8% disagreed. This suggests that while some high school teachers felt the time was adequate, pre-high school teachers were less likely to feel the time allocated was sufficient, with many unsure or neutral about it. The longer time needed for pre-high school activities reflects the additional setup and guidance required for younger students, not their ability. These students benefit from more structured instructions and support, which naturally extends activity time. This suggests that time allocations should be adjusted when planning activities for different age groups.

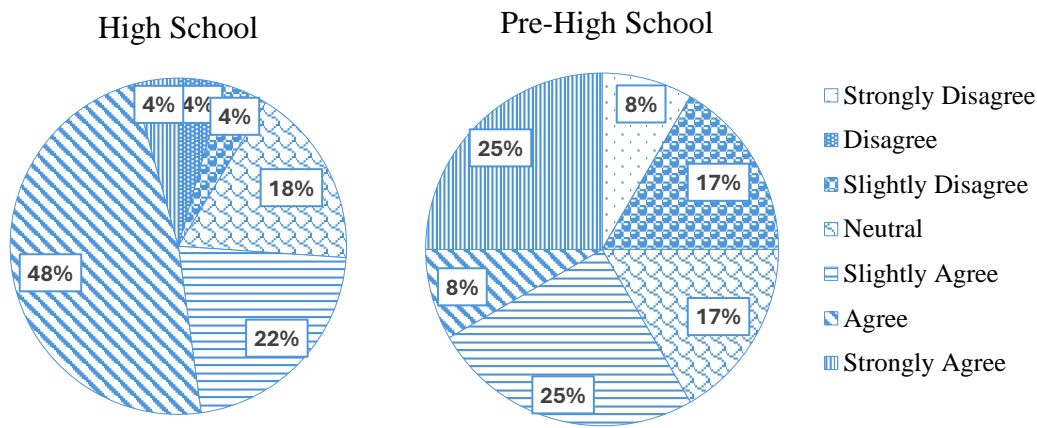

Fig. 5: Percentage responses of the high school and pre-high school teachers to Statement 9

The tenth statement is: "I needed more preparation time to conduct the QuanTime activities/teach quantum concepts well". Fig. 11 (in Appendix) shows that among high school teachers, 18% agreed that they needed more preparation time, a significant portion (59%) of teachers were neutral and 23% disagreed. For pre-high school teachers, 20% agreed, 50% were neutral and 30% disagreed. These responses suggest that the majority of teachers, both at the high school and pre-high school levels, did not feel they needed additional preparation time, indicating that the activities were manageable within the time and resources available. The relatively low percentage of teachers agreeing that they needed more preparation time may imply that the activities were straightforward enough to implement without requiring extensive additional effort as intended.

The eleventh statement is: "I would have liked more support during the implementation of the QuanTime activities/teaching of quantum concepts". Fig. 6 shows that among high school teachers, 13% agreed that they would have liked more support, 65% were neutral and the remaining teachers (22%) disagreed. For pre-high school teachers, 64% were neutral and 36% disagreed, suggesting that most pre-high school teachers did not feel they needed additional support during the implementation of the activities. These responses indicate that while a small percentage of high school teachers felt more support was necessary, most teachers across both levels did not express a strong need for extra assistance.

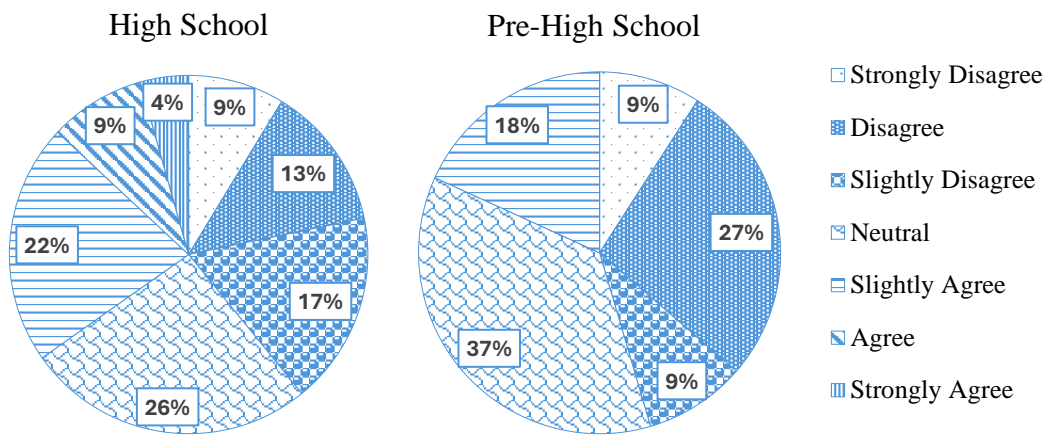

Fig. 6: Percentage responses of the high school and pre-high school teachers to Statement 11

The twelfth statement is: "I would like to do the QuanTime activities/teach quantum again next year". Among high school teachers, 87% agreed, 8% were neutral, and the remaining teachers disagreed. Fig. 12 (in Appendix) shows that for pre-high school teachers, 82% agreed, 9% were neutral, and 9% disagreed. The high percentage of teachers - 87% at the high school level and 82% at the pre-high school level, expressing interest in repeating the QuanTime and other quantum-related activities demonstrates that the teachers may have found them valuable and enjoyable. The desire to continue conducting these activities in their classrooms highlights their positive reception and willingness to integrate them into future lessons. This strong enthusiasm suggests that the activities effectively engaged teachers and likely contributed to their confidence in teaching quantum concepts.

**3.2 Insights from QuanTime Implementation**

We now focus specifically on teacher feedback on QuanTime activities. We analyzed the responses from the teachers who conducted each QuanTime activity in their classrooms, comparing participation at both levels. Table 3 highlights the details of the 8 QuanTime activities, out of more than two dozen provided by the Q-12 partnership program, based on teacher engagement. Each of these activities was conducted by at least one teacher in either pre-high school or high school level. We had two teachers from high school level who conducted two QuanTime activities in their classes. This comparison in Table 3 provides insights into which activities were most frequently implemented and which ones saw less adoption among the teachers who participated in the survey, helping us understand teacher preferences and potential areas for improvement in activity design or appeal.

Table 3: Number of pre-high school (4-8) and high school (9-12) teachers who conducted each QuanTime activity

|  | Pre-HS | HS | Total |
|---|---|---|---|
| Wave-particle duality | 2 | 6 | 8 |
| Electron Transitions | 2 | 4 | 6 |
| Art & Polarization | 4 | 2 | 6 |
| Atomic rainbows | 0 | 2 | 2 |
| TwinTanglement | 1 | 1 | 2 |
| Photoelectric Effect | 0 | 1 | 1 |
| QueueBits | 0 | 1 | 1 |
| Exploring Spectra | 0 | 1 | 1 |

Out of the 8 QuanTime activities that the teachers in our sample used, the most popular was Wave-Particle Duality [63] conducted by 8 teachers, followed by Art & Polarization [64] and Electron Transitions [65], each used by 6 teachers. TwinTanglement [66, 67] and Atomic Rainbows [68] were each conducted by 2 teachers, while Photoelectric Effect [69], QueueBits [69], and Exploring Spectra [70] were each incorporated in their class by 1 teacher. Among high school teachers, Wave-Particle Duality was the most popular, conducted by 6 teachers, followed by Electron Transitions by 4 teachers, and Atomic Rainbows and Art & Polarization, each by 2 teachers. Among pre-high school teachers, Art & Polarization was the most popular, carried out by 4 teachers, followed by Wave-particle duality and Electron Transitions, which were incorporated in their classes by 2 teachers.

We further analyzed each of the 12 statements in Table 2 by examining the distribution of responses from all teachers at both the pre-high school (4-8) and high school (9-12) levels, categorized based on the specific QuanTime activity they conducted in their classrooms. The results of this analysis are presented in Table 4, which provides a detailed breakdown of how teacher responses vary depending on the activity they implemented. This can help understand if engagement with different QuanTime activities influences teachers' perceptions and attitudes. In Table 4, the average of all these statements is also calculated, except for statements 10 and 11, because a higher value for these statements corresponds to a negative response.

Table 4: Distribution of responses from pre-high school (4-8) and high school (9-12) teachers based on the QuanTime activity conducted in their classrooms

|  | Art & Polarization | | Electron Transitions | | Wave-Particle Duality | | Twin-Tanglement | | Atomic Rainbows | |
|---|---|---|---|---|---|---|---|---|---|---|
|  | Pre HS | HS | Pre HS | HS | Pre HS | HS | Pre HS | HS | Pre HS | HS |
| St.1 | 4.8 | 5.5 | 6.5 | 6.5 | 7 | 6.5 | 7 | 7 | - | 6.5 |
| St.2 | 3.5 | 6 | 7 | 6.3 | 4 | 6.4 | 7 | 4 | - | 6 |
| St.3 | 4.8 | 5.5 | 6 | 6.5 | 6 | 6.5 | 7 | 3 | - | 6.5 |
| St.4 | 5.3 | 5.5 | 6.5 | 5.8 | 7 | 6.5 | 7 | 6 | - | 5.5 |
| St.5 | 5 | 6 | 6 | 6.5 | 7 | 6 | 7 | 7 | - | 6.5 |
| St.6 | 5 | 5.5 | 6 | 6.3 | 6 | 6 | 7 | 7 | - | 6.5 |
| St.7 | 3.8 | 4 | 5 | 5.3 | 5 | 5 | 4 | 5 | - | 5.5 |
| St.8 | 3.8 | 4.5 | 6 | 5.5 | 6 | 5.5 | 5 | 6 | - | 6 |
| St.9 | 4.5 | 5.5 | 5.5 | 5.5 | 5 | 5 | 5 | 6 | - | 5 |
| St.12 | 5.5 | 7 | 5.5 | 7 | 7 | 6.5 | 7 | 6 | - | 6.5 |
| AVG | 4.6 | 5.5 | 6 | 6.1 | 6 | 6 | 6.3 | 5.7 | - | 6 |

Table 4 shows that most of these activities have a high average value of 6 which corresponds to 'Agree' in the 7-point Likert scale. Art & Polarization was the only activity that received a relatively low average score of 4.6 at the pre-high school level. It looks like most of the teachers at both high school and pre-high school levels agreed with these statements. This suggests that all these five activities were positively received by the teachers at both levels. The consistently positive reception of all five activities demonstrates that QuanTime activities may be effective in engaging teachers and students across different grade levels.

### 3.3 Teachers' Insights from Open-Ended Questions

From the responses, we found that the most chosen QuanTime activities were Electron Transitions, Wave-Particle Duality, and Art & Polarization. Teachers noted that they selected these activities because they fit well with their existing curriculum, especially in chemistry and physics courses. For example, one teacher noted, "*Electron Transitions best fits in the context of a chemistry course*". Teachers also integrated activities that aligned with students' interests, such as lasers, diffraction gratings, and exploring light properties. These hands-on components generated excitement among students, with one teacher observing, "*They enjoyed just exploring what the light looked like*" and another stating, "*Students loved seeing the differences in glowing when different light sources were used*". Activities that involved visual and interactive elements, such as playing with LEDs and lasers, were particularly engaging, and one teacher noted, "*Getting to play with the lights...they are middle schoolers, so they do not see the fun in the info rather than playing with the items*".

Teachers found Wave-Particle Duality to be a highly engaging and relevant concept for their students. One teacher mentioned that the most exciting part was, *"Collecting results and plotting graphs because they worked in groups and the visuals solidified understanding of wave-*

*particle duality. The concept of duality was surprising and fun".* This activity allowed students to explore the intriguing nature of light behaving both as a wave and a particle. Another teacher noted that the exciting part of the activity was, *"The questions they had about their ideas of what a particle versus a wave was".* Overall, Wave-Particle Duality proved to be an impactful activity, sparking curiosity and encouraging critical thinking about the nature of light and quantum phenomena.

Unique activities like TwinTanglement intrigued students, as one teacher mentioned, *"Students were excited about the concepts of correlation between the twins. Many students could instinctively advance to a higher level".* This quote highlights the students' enthusiasm to get introduced to the fascinating idea of quantum entanglement through this engaging activity. Overall, the varied responses indicate a strong preference for activities that are visually appealing and interactive, aligning with the teachers' goals of making quantum concepts accessible and engaging for students.

Most of the teachers felt that the QuanTime activities aligned well with their quantum concept lessons. Several highlighted specific instances where the activities complemented their existing curriculum, such as during units on waves, electron transitions, and astronomy. One teacher noted, "*The Electron Transitions fit in well with the time and lessons I was teaching at the time as we were talking about the gain and loss of electrons based on the conditions.*" Another teacher stated, "*It fit very well with our mechanical Waves unit.*" Some noted using additional resources like "Quantum for All" [62], which added depth to their teaching. A few respondents did not provide any feedback on whether these activities aligned well with their quantum classes (left the question blank). Overall, the majority of the teachers confirmed that the quantum activities fit well and enriched the learning experience for their students.

Most teachers reported feeling confident in answering student questions while engaging in QuanTime activities, with many indicating they did not encounter questions from their students they couldn't address. For instance, one teacher mentioned, "*No. I am pretty confident with this material*" while another stated, "*Since I have worked with Quantum4All, I did not have a student question that I could not answer*". Even though we did not explicitly ask teachers about their prior experience with teaching quantum concepts, a few voluntarily mentioned having participated in quantum workshops on their own. Thus, some of the teachers, particularly those who mentioned having prior experience in quantum education, felt well-prepared to handle inquiries from their students. However, a few teachers expressed that some topics, such as quantum computing, raised questions they weren't confident answering, with one teacher saying, "*There were questions about quantum computing that I didn't feel confident answering. How do quantum computers work?*" Moreover, some teachers mentioned that extra resources would have been helpful, as one teacher noted, "*Extra sources would have been great, I would have been able to direct those students to a source that they could have used to look up their own questions.*" While some teachers felt completely comfortable, others reflected that some questions were challenging, with one sharing, "*Yes, some questions were difficult and yes I would have liked help for them.*" One teacher also noted that, although questions sometimes arose, "*Students knew I couldn't help them, so they*

*figured it out by themselves and had fun!*" Overall, the feedback suggests that while most teachers felt equipped to answer questions, there is a need for further resources to assist with more complex topics in quantum science.

We also asked the educators if they shared the QuanTime activity with other educators. We find that most of the teachers did not share the QuanTime activity with others. There were 11 respondents who mentioned that they shared QuanTime activities with others. 2 respondents clearly stated that no other teachers implemented the activities whereas 2 other respondents mentioned that one person implemented QuanTime activities apart from them. A few mentioned sharing it with pre-service teachers or planning to share it next year. One respondent also had a colleague observe the activity in class. Most teachers reported using QuanTime activities either as part of their regular curriculum or as a connected but fun add-on. About half integrated it directly into existing lessons or connected it with prior topics covered, while the rest treated it as an unconnected addition to their teaching to get students to learn quantum concepts while having fun. Several noted plans to incorporate it more formally into their curriculum in the future, emphasizing its potential for enhancing student engagement and expanding their understanding of quantum concepts.

In addition to these insights, several educators shared broader suggestions and feedback for improving the QuanTime activities. One educator highlighted that the resources could be valuable not only for K-12 students but also for first-year physics undergraduates noting, "*I know K-12 is the focus, but these resources are valuable for first-year physics undergraduates and non-math/science undergraduates. I've also used them in summer STEM camp (in connection with Quantum for All). I'd like to encourage first-year physics and other science students with slight modifications of these activities. They hear "quantum" and want to walk away, but Quantum is FUN! They need to know that and the importance of it!!*". Teachers also expressed interest in continuing to use the materials, with one expressing, "*I would love to do this again and try another kit*" and another saying, *"I hope to pick two [activities] next year*". Another educator mentioned incorporating the activities into their IDK (I Don't Know What to Do) materials, providing ongoing opportunities for students to explore quantum concepts, "*I have now included the kit in my IDK (I Don't Know What to Do) materials. This is going to be a continuous option for my students to explore - especially after our wave and light experiments in the classroom. Depending on the passions of the students, I am excited to have this as an option for discourse"*. A few suggested making the materials more accessible by adding grade-level differentiation, "*Having the presentation slides coded in some manner for grade level differentiation would be helpful (green frame for elementary, yellow for middle, red for H.S. [high school] for example) for scaffolding/presenting*". Overall, the resources were praised for being easy to use, with teachers noting, "*The lesson was well designed and the materials simple to acquire and use. I will show more attention to presenting it earlier in the year before my students quit*" while others asked for clearer instructions, as one teacher remarked, "*Send better instructions*". Despite minor challenges, there was strong encouragement to continue the program, with teachers expressing appreciation like, "*Thanks for your efforts*", and "*Keep going with this! It is accessible and interesting*".

## 4. Conclusion and Discussion

Teacher feedback from this study suggests that QuanTime and other quantum-related activities may serve as a promising approach for introducing quantum concepts in K–12 settings and helping students begin to engage with the quantum world. The feedback from teachers provides essential insights into the most engaging aspects of these activities and suggests ways to improve others to ensure they are equally captivating. Since the materials are openly available (at least the examples in this paper), they can serve to model what an easily implementable activity should look like. If at some future time it turned out that the activities are not particularly efficient for learning, their structure and organization can still help with the structure and organization of new activities. By refining these activities based on teacher input, it is possible to create a more comprehensive and impactful educational experience for students.

This research highlights the value of hands-on, interactive learning in quantum education, as reflected in the positive responses from teachers regarding several QuanTime and other quantum related activities. In particular, teachers identified visually engaging and interactive activities—such as Art and Polarization, Wave-Particle Duality, and Electron Transitions—as especially effective in capturing student interest and facilitating classroom discussions. These activities were among the most frequently used and received high ratings in our survey, along with favorable open-ended comments about how they helped make complex quantum ideas more tangible and accessible. While we cannot draw conclusions about direct student learning outcomes, the consistency of teacher feedback suggests that such approaches are perceived as useful tools for introducing quantum concepts in K–12 classrooms. This preference for interactive and visually appealing activities emphasizes the importance of bridging the gap between theoretical knowledge and practical understanding in science education.

Most teachers reported that students genuinely enjoyed participating in the QuanTime activities, which is a highly encouraging outcome. Enjoyment plays a key role in sparking curiosity and serves as an important gateway to developing interest in new topics. In the context of quantum science, such positive and engaging experiences can lay the groundwork for future interest and exploration. While some students may not develop a strong or lasting interest right away, this is expected and consistent with Hidi and Renninger's model of interest development, which emphasizes that triggered situational interest—like enjoyment from a novel activity—is the first step. With continued exposure and support, this initial enjoyment has the potential to evolve into deeper engagement with quantum concepts over time.

We also find that most teachers expressed confidence in conducting the QuanTime activities, with many reporting that the allocated time and support were sufficient for effective implementation. This suggests that the activities were not only feasible within the curriculum constraints but also contributed to teachers feeling more prepared and capable in teaching quantum concepts. One possible reason for this positive response is that QuanTime activities supplement rather than replace core curriculum content. Since they are not tied to assessments, teachers and students may feel more comfortable exploring them in a low-pressure, curiosity-driven environment. However, we can infer from the survey responses that more scaffolding, and preparation would be beneficial. Most teachers at both high school and pre-high school levels were enthusiastic about continuing to use these activities in the future. The open-ended responses provided further insight into the aspects of these activities that students enjoyed most, such as their interpretations of particles versus waves in Wave-Particle Duality, observing variations in glowing with different light sources in Electron Transitions, and exploring the concept of correlation between twins in TwinTanglement. These findings underscore that the teachers perceived the activities as effective in engaging students and supporting classroom conversations about complex quantum topics.

An important limitation of this study is the low response rate—only 36 out of the 132 invited teachers completed the survey. While this was noted previously, it is important to further consider the potential implications of this limited participation. In particular, the possibility of non-response bias must be acknowledged. It remains unclear why the majority of teachers chose not to respond to our survey. It is possible that non-respondents did not think it was necessary to respond to our survey due to the time required to complete it. It is also possible that they were less engaged with the QuanTime materials, opted not to implement the activities, or encountered barriers—such as time constraints, lack of support, or limited confidence—that prevented participation. If so, their perspectives could have offered important insights into the challenges educators face when attempting to incorporate quantum-related content into their teaching. As a result, the views represented in this study may reflect those of teachers who were more motivated, confident, or already interested in quantum science, which may limit the generalizability of our findings. Acknowledging this potential bias is essential for interpreting the results with appropriate caution and for guiding future efforts to support a broader and more representative group of educators in quantum education initiatives.

Some of the QuanTime activities were not used by the teachers at either level in our sample. The reasons for this are unclear; it would be interesting to explore teachers' perspectives on why they chose certain activities in future studies. It is also important to note that there was a lack of computer science teachers in our sample, and the QuanTime activities about quantum games specifically focused on concepts and modalities for those classrooms. Further research could help identify the specific aspects of the most used activities that made them appealing to the teachers by interviewing the teachers who used them individually or in a small focus group.

Another limitation of this study is that we did not collect additional data to clarify how teachers interpreted the survey questions or how students engaged with the quantum-related activities in their classrooms. Teachers responded to the survey items based on their own interpretations, without opportunities for follow-up explanations or elaboration. As a result, there may be variability in how key terms or prompts were understood, which could affect the consistency and reliability of the responses. Furthermore, since we did not gather direct information about student engagement or learning outcomes, we are unable to assess how the activities were received by students or the extent to which they were implemented as intended. These gaps limit our ability to draw deeper conclusions about the classroom impact of QuanTime and highlight the need for future studies that incorporate more detailed qualitative data from both teachers and students.

We also observed limited dissemination of QuanTime activities suggesting that the resources may not be distributed within the broader STEM teaching community. One possible reason could be that teachers may hesitate to promote quantum-related activities if their colleagues themselves lack interest or confidence in the subject. Quantum physics is a complex and relatively unfamiliar topic for many K–12 educators, and without foundational knowledge or motivation, teachers may be reluctant to engage with or share such resources. This reluctance could stem from colleagues perceiving quantum education as "extra" work beyond already demanding curricula or feeling unprepared to incorporate it effectively. Increasing teacher content knowledge and self-efficacy may not only improve their willingness to implement quantum activities but also foster a culture of enthusiasm that encourages resource sharing and collaborative professional learning.

It is also important to note that teachers who have had positive or empowering experiences with quantum content may be more likely to approach these activities with confidence and excitement, which in turn can influence how students perceive and engage with the material. From the perspective of Expectancy-Value Theory (EVT), a teacher's own expectancy for success and perceived value of the content can shape their instructional approach and the motivational climate they create for students. Similarly, Social Cognitive Career Theory (SCCT) highlights how contextual influences, such as teachers' own learning histories and outcome expectations, may affect their sense of efficacy and the goals they set for their students. Teachers who view quantum as both important and enjoyable are more likely to convey this message to students, potentially helping to shift student attitudes from avoidance to curiosity. This points to the importance of supporting teachers' development in quantum literacy—not only to enhance their content knowledge, but also to positively shape their attitudes and instructional enthusiasm, which can have cascading effects on student engagement and interest development.

Moreover, this study did not assess student understanding or learning outcomes, nor did it directly measure student interest in quantum concepts beyond teacher reports. As such, conclusions about the impact of QuanTime on student learning or the development of sustained interest must be drawn with caution. Theoretical frameworks, such as Hidi and Renninger's interest development model, suggest that sparking initial situational interest is only a first step; cultivating deeper and lasting interest requires ongoing support and repeated engagement, aspects not explored here. A potential future direction could involve conducting a survey of their students to gain insights from student perspective directly into the different aspects of QuanTime activities instead of through their teachers as in this investigation. This approach would provide additional insight into the aspects of the activities that are effective and the areas that may need improvement from students' point of view, which may or may not align with their teachers' view discussed here. These insights could then be used as a guide to refine and enhance the activities including those not used frequently by the teachers in our sample. It would also be valuable to have teacher and student feedback on QuanTime activities from across different types of schools including those in countries other than the US. Collecting data from a more diverse teacher population would further enhance our understanding of the broader applicability of QuanTime and support its continued refinement.

Despite the limitations, our study provides timely and valuable insight into how K–12 teachers engage with quantum-related activities in their classes. Furthermore, by examining the responses of both pre-high school and high school teachers, we shed light on how quantum concepts can be meaningfully introduced at varying developmental stages. Our findings demonstrate that teachers generally found QuanTime and other quantum-related activities to be engaging, easy to implement, and aligned with their curricular goals, with many reporting high levels of student enjoyment and interest. Importantly, this research brings attention to the practical and pedagogical potential of integrating quantum science into K–12 education. It shows that when supported with thoughtfully designed materials, teachers can successfully introduce abstract quantum concepts in engaging ways to their students. The QuanTime initiative serves as a useful model for how national partnerships can foster innovation in science education, especially in these emerging fields. In conclusion, this study contributes to both research and practice by illuminating how early exposure to quantum ideas can support broader goals of scientific literacy, while also affirming the readiness and enthusiasm of teachers to take up this important work.

## ACKNOWLEDGEMENT

We sincerely thank the National Science Foundation (NSF) for supporting our research through awards NSF-PHY-2309260 and NSF DMR-2039745. We also gratefully acknowledge Jen Palmer, a staff member in Franklin's group, for their invaluable contributions to the creation and continued support of QuanTime. We acknowledge and extend our gratitude to all the creators, funders, and organizers of the QuanTime activities. We also appreciate and acknowledge the teachers who participated in this program.

APPENDIX

Statement 1

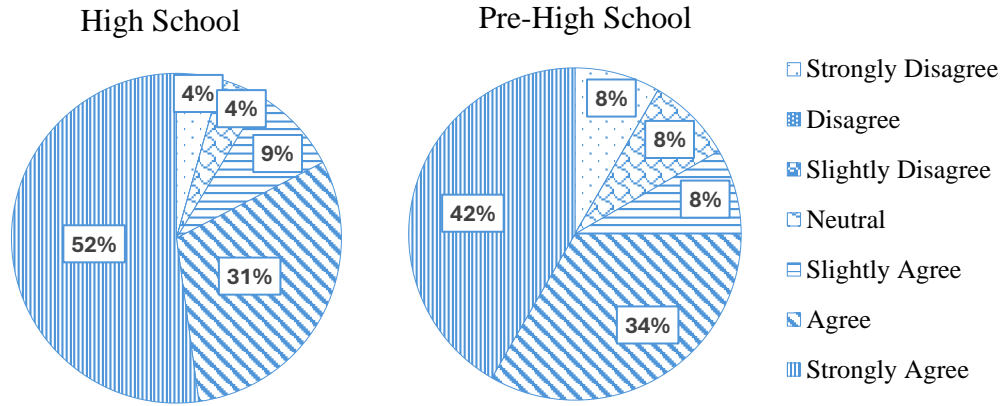

Fig. 7: Percentage responses of the high school and pre-high school teachers to Statement 1

Statement 6

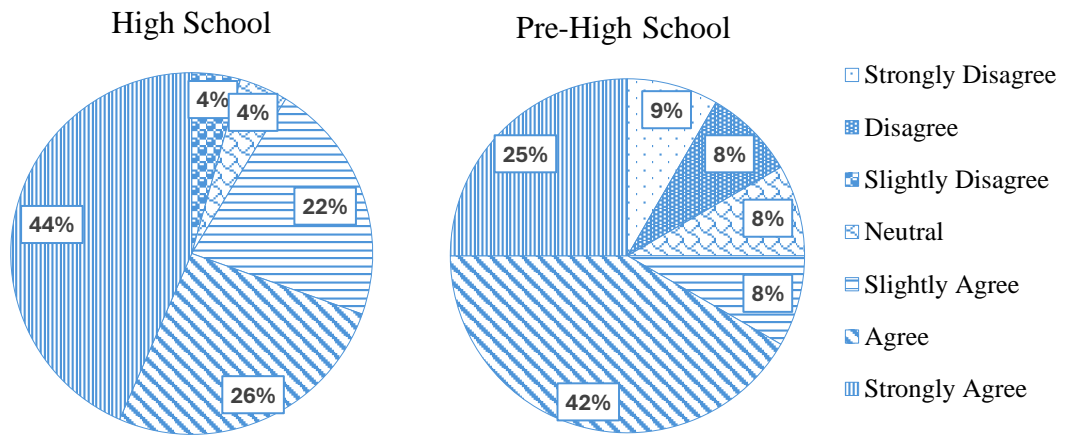

Fig. 8: Percentage responses of the high school and pre-high school teachers to Statement 6

Statement 7

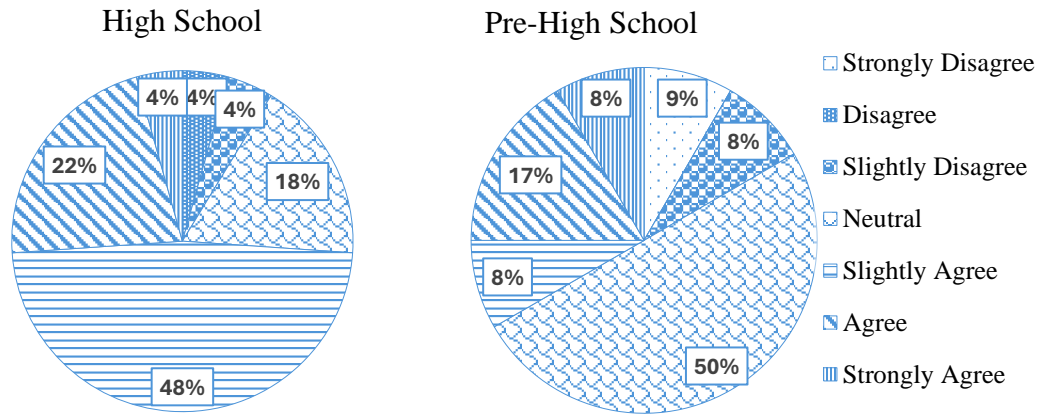

Fig. 9: Percentage responses of the high school and pre-high school teachers to Statement 7

Statement 8

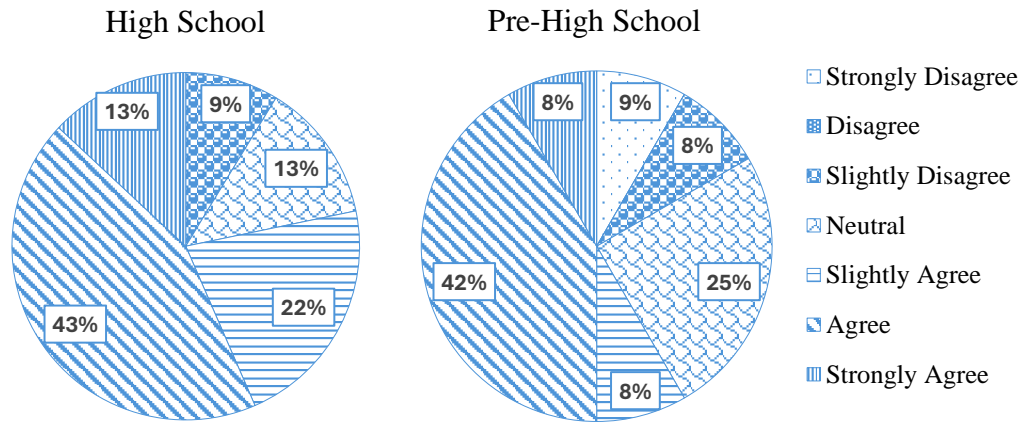

Fig. 10: Percentage responses of the high school and pre-high school teachers to Statement 8

Statement 10

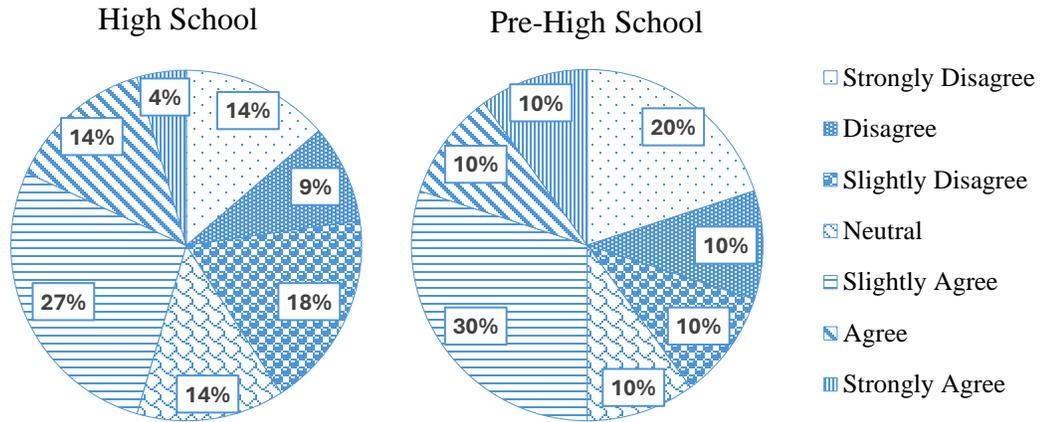

Fig. 11: Percentage responses of the high school and pre-high school teachers to Statement 10

Statement 12

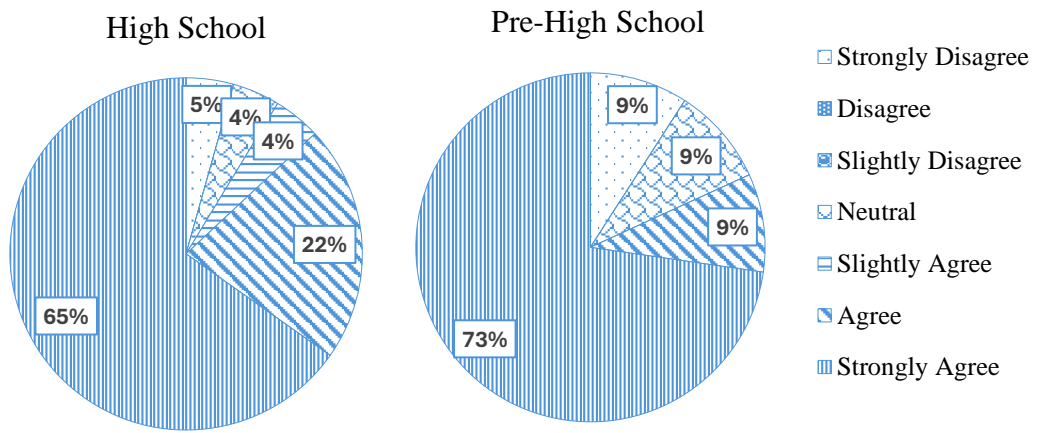

Fig. 12: Percentage responses of the high school and pre-high school teachers to Statement 12